\documentclass[
    platex, 
    aps,
    prl,
    twocolumn,
    superscriptaddress,
    amsmath,amssymb,
]{revtex4-2}

\usepackage[dvipdfmx]{graphicx} 
\usepackage{bm}
\usepackage{xr} 
\usepackage[breaklinks=true]{hyperref}

\hypersetup{%
  colorlinks=true,%
  linkcolor=blue,%
  citecolor=blue,%
  urlcolor=black,%
}

\begin{document}


\title{
    Programming Spintronic Reservoir Computing
}


\author{Yuichiro Terasaki}
\email[]{terasaki@isi.imi.i.u-tokyo.ac.jp}
\affiliation{Graduate School of Information Science and Technology, The University of Tokyo, Tokyo 113-8656, Japan}

\author{Yusuke Imai}
\email[]{imai@isi.imi.i.u-tokyo.ac.jp}
\affiliation{Graduate School of Information Science and Technology, The University of Tokyo, Tokyo 113-8656, Japan}

\author{Jason Z. Kim}
\email[]{jk2557@cornell.edu}
\affiliation{Department of Physics, Cornell University, Ithaca, NY, USA}

\author{Kohei Nakajima}
\email[]{k-nakajima@isi.imi.i.u-tokyo.ac.jp}
\affiliation{Graduate School of Information Science and Technology, The University of Tokyo, Tokyo 113-8656, Japan}
\affiliation{Next Generation Artificial Intelligence Research Center, The University of Tokyo, Tokyo 113-8656, Japan}

\date{\today}

\begin{abstract}
    We present a programming framework for a spintronic reservoir computer (RC) that maps prescribed input-output relationships directly onto the readout layer, bypassing conventional data-driven black-box approaches.
    Our spintronic RC is based on magnetoresistive random-access memory and exploits magnetization dynamics for computation.
    We introduce a general metric that quantifies the system's programmability and reveals how the governing equations and system parameters constrain the class of realizable functions.
    We then construct externally controllable readout layers by exploiting the explicit parameter dependence of the prescribed equations.
    This metric and construction enable programming explicit functions on the spintronic RC, indicating a potential route to in-memory computing.
    Our demonstrations include neural-network emulation, bifurcation embedding, and a Newton solver for fifth-order algebraic equations.
    In addition, we prove the universal approximation property of the spintronic RC in the limit of infinite system size and input duration, and show its consistency with programmability.
\end{abstract}


\maketitle


\textit{Introduction}.---
Recent developments in physical computing and physical neural networks have explored unconventional paradigms for information processing, offering pathways to surpass traditional performance limits (e.g.,~processing speed or cost per operation) and revealing the computational capabilities of systems once regarded primarily as materials
\cite{
    nakajima2020physical,
    nakajima2021reservoir, wright2022deep, nakajima2022physical, jaeger2023toward, momeni2025training, wang2026embodying}.
In particular, physical reservoir computing (PRC) enables us to extract usable computational resources from the high-dimensional nonlinear responses inherent to material reservoirs simply by designing the observation, or readout, method
\cite{nakajima2020physical,nakajima2021reservoir}.
Nevertheless, optimal PRC readouts have usually been discovered through data-driven black-box approaches.

\begin{figure}
    \includegraphics[width=\columnwidth]{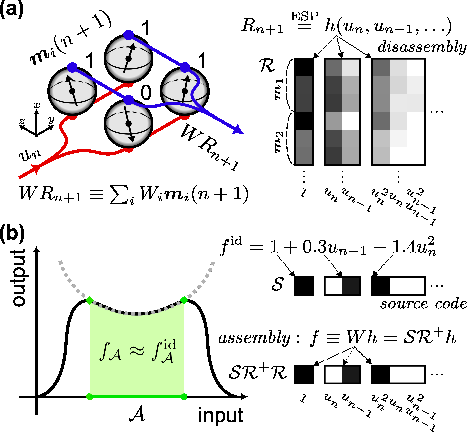}
    \caption{
        Programming reservoir computing $f$ using spintronic dynamics in the memory storage.
        (a)
        The readout is modeled as a linear combination of the individual magnetization states.
        Each magnetization state $\bm{m}_i$ is decomposed onto analytic bases of the input sequence $\{u_n, u_{n-1}, \dots \}$, yielding the \textit{disassembly} matrix $\mathcal{R}$.
        (b)
        The ideal computation $f^{\rm id}$ is also decomposed using the same analytic bases, yielding the \textit{source code} $\mathcal{S}$.
        The RC system approximates $f^{\rm id}$ within an effective input region $\mathcal{A}$ via $W = \mathcal{S}\mathcal{R}^{+}$, called the \textit{assembly} process.
    }
    \label{fig:PSRC}
\end{figure}


In this Letter, we establish a scheme for embedding functions with explicit formulas into physical reservoirs, enabling computations to be translated across different physical substrates while exploiting the assets of each.
For example, photonic reservoirs are naturally suited to solving optimization problems due to their high processing speed. Spintronic reservoirs also have fast time constants and robustness in radioactive environments \cite{akashi2022coupled}.
This scheme differs from learning, which seeks to generalize from data when the target function is unknown beforehand.
In contrast, our method is \textit{programming}: the reservoir dynamics are \textit{disassembled} into interpretable bases, \textit{source code} is written from a target function, and the readout is \textit{assembled} via the pseudo-inverse of the disassembly matrix (Fig.~\ref{fig:PSRC}).
Prior work has shown that programming reservoir computing (RC) brings out the expressivity of random neural networks and enables practical operations (e.g., Fourier transformation, virtualization of small RC systems, and \textit{Pong} gameplay) \cite{kim2023neural}.
Here, we apply the RC programming framework to physical systems, namely in the realm of PRC.
According to this extension, multiple novel schemes and investigations should be introduced into the approach.
First, unlike the conventional machine learning network, expressivity of physical systems are unknown in general.
To evaluate PRC expressivity, we introduce a programmability metric and discuss the universal approximation property (UAP) for spintronic RC systems in the limit of infinite system size and input duration.
This metric facilitates the analysis of non-universal physical systems,
including spintronic RC systems operating with finite sizes and input durations.
We demonstrate our framework by implementing in-memory computing on magnetoresistive random-access memory (MRAM), in which the hardware functions simultaneously as memory and processor \cite{sebastian2020memory,sun2023full,marrows2024neuromorphic}.
We also evaluate the robustness of programmed RC against the thermal noise inherent to the spintronic systems.

We require the reservoir system to satisfy two primary conditions:
(i) the reservoir response $R_{n+1}$ can be described as a function of input timeseries $\{u_n, u_{n-1}, \dots \}$:
\begin{equation}
    \label{eq:ESP}
    R_{n+1} = h(u_n, u_{n-1}, \dots),
\end{equation}
and (ii) the readout is given by a linear combination of reservoir states [Fig.~\ref{fig:PSRC}(a)].
Condition (i) is related to the echo-state property (ESP) in the context of RC \cite{jaeger2007echo}.
Programming PRC then determines the readout matrix $W$ in an interpretable way to align the PRC output $f \equiv Wh$ with that of the target equation $f^{\rm id}$ over the effective input region $\mathcal{A}$ [Fig.~\ref{fig:PSRC}(b)] \cite{komatsu2024algebraic}.

\textit{Disassembly, source code, and assembly}.---
Because the input $u_n$ is interpretable in most computations, we decompose the reservoir's internal states $R_{n+1}$ onto analytic bases of the input sequence $\{u_n, u_{n-1}, \dots \}$.
This transformation is termed \textit{disassembly} because the system's physical configurations are converted into interpretable computational components.
We characterize this process by partially differentiating Eq.~\eqref{eq:ESP} with respect to a specific combination of input-history variables (e.g.,~$\{u_n, u_{n-1}\}$) under the steady-state input assumption $u_n = u_{n-1} = \cdots = p \,\, (p \in \mathcal{A})$.
In general, we consider all combinations of $k$th-order partial derivatives with respect to the latest $d+1$ input-history terms $\{u_{n}, \ldots, u_{n-d}\}$ at $u_n = p$.
We denote this series of operations by $\partial^{k_d}_p$, where $k_d$ is the multi-index for differentiation with $|k_d| = k$.
The total number of partial-derivative combinations is $\binom{d_{in} \cdot (d+1) + k - 1}{k}$ when the input dimension is $d_{in}$.
We call the resulting matrix $\mathcal{R}^{k_d}_p \equiv \partial^{k_d}_p h$ the \textit{disassembly matrix}.
An approximation $f \equiv Wh \approx f^{\rm id}$ based on the decomposition $\partial^{k_d}_p$ determines the unique readout $W$ via the Moore-Penrose pseudo-inverse of $\mathcal{R}^{k_d}_p$:
\begin{equation}
    \label{eq:programming1}
    W
    \equiv
    \mathcal{S}^{k_d}_p
    \left(
        \mathcal{R}^{k_d}_p
    \right)^{+}
    \quad
    \text{where}
    \quad
    \mathcal{S}^{k_d}_p \equiv \partial^{k_d}_p f^{\rm id}.
\end{equation}
The matrix $\mathcal{S}^{k_d}_p$ is termed the \textit{source code} and represents the decomposition of the target function $f^{\rm id}$ into interpretable computational components along $\partial^{k_d}_p$.
Accordingly, \textit{assembly} refers to the process of mapping the source code $\mathcal{S}^{k_d}_p$ onto the readout matrix $W$ via the pseudo-inverse of the disassembly matrix $\mathcal{R}^{k_d}_p$.
Our framework permits an arbitrary number of decompositions $\{\partial^{k_d}_p\}$ with arbitrary parameters $k$, $d$, and $p$.
Let $\mathcal{R}$ and $\mathcal{S}$ denote the concatenations of the disassembly matrices $\left[\mathcal{R}^{{k'}_{d'}}_{p'}; \mathcal{R}^{{k''}_{d''}}_{p''}; \cdots \right]$ and source codes $\left[\mathcal{S}^{{k'}_{d'}}_{p'}; \mathcal{S}^{{k''}_{d''}}_{p''}; \cdots \right]$, respectively, over a set of decompositions $\{\partial^{k_d}_p\}$.
Then, the source code $\mathcal{S}$ is assembled into the readout matrix via $W=\mathcal{S}\mathcal{R}^{+}$.
We note that choosing a different value of $p$ does not imply decomposition with respect to different input variables; the matrix $\mathcal{R}$ simply concatenates different representations of the RC's expressivity for the same set of input variables.

\textit{Programmability}.---
Since the analytic bases selected for programming are not necessarily supported by the underlying physical hardware, a general metric is needed to assess which functional components of the reservoir system are programmable.
To define this quantity, we express the decomposition of the RC output  $W\mathcal{R}$ as a projection of the source code $\mathcal{S}$ via a specific symmetric matrix, $\mathcal{R}^{+} \mathcal{R} = \Theta^{+}\Theta$, where $\Theta \equiv \mathcal{R}^\top\mathcal{R}$:
\begin{equation}
    \label{eq:Theta+Theta}
    W
    \mathcal{R}
    =
    \mathcal{S}
    \mathcal{R}^{+}
    \mathcal{R}
    =
    \mathcal{S}
    \left(
        \mathcal{R}^\top
        \mathcal{R}
    \right)^{+}
    \mathcal{R}^\top
    \mathcal{R}
    =
    \mathcal{S}
    \Theta^{+}
    \Theta
    .
\end{equation}
This formulation using $\Theta$ allows us to take the infinite-system-size limit of programmability [see Sec.~I in the Supplementary Material (SM)].
Consequently, the diagonality of $\Theta^{+}\Theta$ quantifies the consistency among different decompositions, indicating which sets of analytic bases can be programmed simultaneously in the system.
We call $\Theta^{+}\Theta$ the \textit{programmability matrix}.
In numerical calculations, the pseudo-inverse of $\Theta\equiv\mathcal{R}^{\top}\mathcal{R}$ may overlook tiny singular values of $\mathcal{R}$ that are close to machine epsilon.
Although a designer must strategically choose analytic bases based on programmability to approximate $f^{\rm id}$ in the effective input space $\mathcal{A}$, the effects of system parameters on overall programmability must also be considered.

\begin{figure}
    \includegraphics[width=\columnwidth]{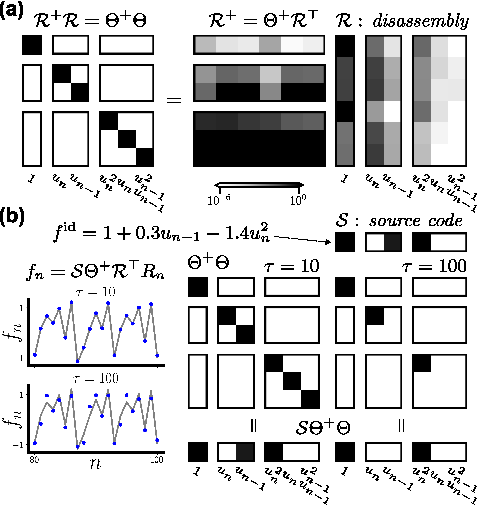}
    \caption{
        Programmability of the spintronic RC system with $b=0.1$.
        (a) Programmability matrix $\Theta^{+}\Theta$ under $k$th-order analytic bases of $\{u_{n}, u_{n-1}\}$ with $k=0,1,2$ and input duration $\tau=10$.
        The color scale represents the absolute value of each matrix element.
        The diagonality of $\Theta^{+}\Theta$ indicates that each analytic basis can be independently controlled through the readout $W$.
        (b) One-step-ahead prediction of chaotic dynamics programmed under the same analytic bases for $\tau=10$ and $100$.
        Degeneracy in $\Theta^{+}\Theta$ deforms the programmed function, leading to the mismatch between the programmed and target trajectories for $\tau=100$.
    }
    \label{fig:Theta+Theta}
\end{figure}

\begin{figure*}
    \includegraphics[width=\textwidth]{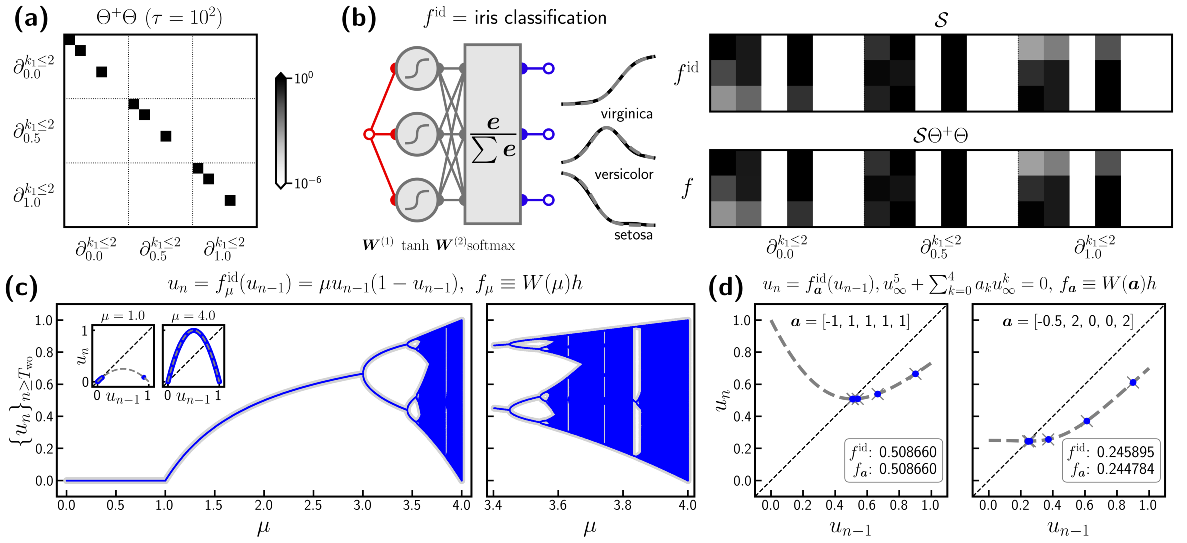}
    \caption{
        Programming the spintronic RC using partial derivatives $\{\partial^{k_1\leq2}_{0}, \partial^{k_1\leq2}_{0.5}, \partial^{k_1\leq2}_{1}\}$ with $\tau=10^2$ and $b=1$.
        (a) Diagonality of $\Theta^{+}\Theta$. The nonzero diagonal elements corresponding to the current input $u_n$ offer programmability for computing $f^{\rm id}(u_n)$. 
        (b) Emulation of a trained neural network for Iris data classification.
        (c) Embedding bifurcations of the dynamical system $u_n = \mu u_{n-1} (1-u_{n-1})$ via the readout $W(\mu)$, which is modulated by the control signal $\mu$.
        (d) Solving the fifth-order algebraic equation $x^5 + \sum_{k=0}^{4} a_k x^k = 0$ using the readout $W(\bm{a})$ with control signals $\bm{a}=[a_0,a_1,a_2,a_3,a_4]^{\top}$.
    }
    \label{fig:programming}
\end{figure*}

\textit{Programming spintronic reservoir computing}.---
Here, to define the underlying physics of the RC system, we model MRAM using the Landau--Lifshitz--Gilbert equation \cite{gilbert2004phenomenological}, which is consistent with recent research on spintronic computing 
\cite{
    chen2025spintronic, akashi2020input, imai2022noise, imai2025gradient, kurebayashi2026metrics}.
We describe the memory states as the $x$-component orientations of the spatially-multiplexed magnetization $\bm{m}_{i}$ [Fig.~\ref{fig:PSRC}(a)]:
\begin{equation}
    \label{eq:LLGReservoir_1}
    \begin{gathered}
        \frac{d\bm{m}_{i}}{dt}
        =
        -
        \frac{|\gamma|}{1+\alpha_i^2}
        \left\{
            \bm{m}_{i} \times \bm{H}_{i}
            +
            \alpha_i \bm{m}_{i} \times
            \left(
                \bm{m}_{i} \times \bm{H}_{i}
            \right)
        \right\}
         ,
        \\
        R_{n}^{(i)} = \bm{m}_{i}(n\tau),
        \quad
        i = 1,\ldots,N,
    \end{gathered}
\end{equation}
where $\tau$ is the input duration, $N$ is the number of nodes, $\gamma$ is the gyromagnetic ratio, $\alpha_i$ is the Gilbert damping factor, and $\bm{H}_{i}$ is the effective magnetic field (comprising the external input $u_n$ and internal interactions).
Other driving schemes near a fixed point can also be considered, including subcritical spin-torque injection and voltage-induced modulation of magnetic anisotropy.
Assuming a fixed input $u_n$ during the interval $n\tau \leq t < (n+1)\tau$, we obtain the following analytic expression for $\bm{m}_{i}[n] \equiv \bm{m}_{i}(n\tau)$ (see Sec.~II in the SM for the derivation):
\begin{equation}
    \label{eq:LLGReservoir_2}
    \begin{aligned}
        \bm{m}_{i}&[n+1]
        =
        \\
        &
        \frac{\sin \phi_n^{(i)}}{\sin \varphi_n^{(i)}}
        \left\{
            \cos \theta_n^{(i)} \bm{m}_{i}[n]
            +
            \sin \theta_n^{(i)} \frac{\bm{H}_{i}(u_n)}{\|\bm{H}_{i}(u_n)\|} \times \bm{m}_{i}[n]
        \right\}
        \\
        &
        +
        \left(
            \cos \phi_n^{(i)}
            -
            \frac{\sin\phi_n^{(i)}\cos \theta_n^{(i)}}{\tan \varphi_n^{(i)}}
        \right)
        \frac{\bm{H}_{i}(u_n)}{\|\bm{H}_{i}(u_n)\|}
       ,
    \end{aligned}
\end{equation}
where
\begin{equation}
    \label{eq:LLGangles}
    \begin{gathered}
        \bm{H}_{i}(t) \equiv \bm{H}_{i}(u_n)
        \,\,
        (n\tau \leq t < (n+1)\tau),
        \\
        \varphi_n^{(i)}
        \equiv
        \arccos
        \frac{\bm{H}_{i}(u_n) \cdot \bm{m}_{i}[n]}{\|\bm{H}_{i}(u_n)\|},
        \,\,
        \theta_n^{(i)}
        \equiv
        \frac{|\gamma|\|\bm{H}_{i}(u_n)\|}{1+\alpha_i^2}\tau,
        \\
        \phi_n^{(i)}
        \equiv
        2 \arctan
        \left(
            e^{-\alpha_i \theta_n^{(i)}}
            \tan \frac{\varphi_n^{(i)}}{2}
        \right)
        .
    \end{gathered}
\end{equation}
Given the existence of a stable fixed point at $\frac{\bm{H}_i}{\|\bm{H}_i\|}$ (the unit vector along the external field) for $\alpha_i > 0$, the ESP [condition (i)] of the RC system is guaranteed (see Sec.~III~A in the SM, which includes Ref.~\cite{grigoryeva2019differentiable}).
We assume linear contributions of $u_n$ to the effective field $\bm{H}_{i}$:
\begin{equation}
    \label{eq:RCinput}
    \bm{H}_{i}(u_n) \equiv \bm{H}_{i}^{(0)} + \bm{H}_{i}^{(RC)},
    \,\,
    \bm{H}_{i}^{(RC)} \equiv W^{in}_i u_n + b^{in}_i,
\end{equation}
where the matrices $\{W^{in}_i\}$ and $\{b^{in}_i\}$ are randomly drawn from $\mathrm{i.i.d.}$ uniform distributions over $[-b,b]$. 
Here, $\bm{H}_{i}^{(0)}$ represents the ``pure'' memory states in the absence of RC input.
While standard MRAM relies on a coarse-grained binary readout (parallel or antiparallel magnetization), the RC system exploits the underlying continuous magnetization dynamics for computation.
In the numerical simulations, we set $|\gamma| = 1, \alpha_i = 1, b \in \{0.1,1\}, \bm{H}_{i}^{(0)} = \pm [1, 0, 0]^{\top}$, and $N={32}^2$, with normalized magnetization ($\|\bm{m}_i\| \equiv 1$).
Note that $|\gamma|$ and $\|\bm{H}_i\|$ effectively scale the time axis without altering the fundamental dynamics.
Figure~\ref{fig:Theta+Theta} illustrates programmability for these physical configurations.
Increasing the input duration $\tau$ reduces the influence of past inputs as per the ESP, which is characterized by zero entries in $\Theta^+\Theta$ for analytic bases containing $u_{n-d} \,\, (d \geq 1)$ at large $\tau$ [Fig.~\ref{fig:Theta+Theta}(b)].
Off-diagonal entries in $\Theta^+\Theta$ imply that incompatible decompositions exist for this system, which also contributes to its nonprogrammability (see Sec.~III~B in the SM).

\begin{figure}
    \includegraphics[width=\columnwidth]{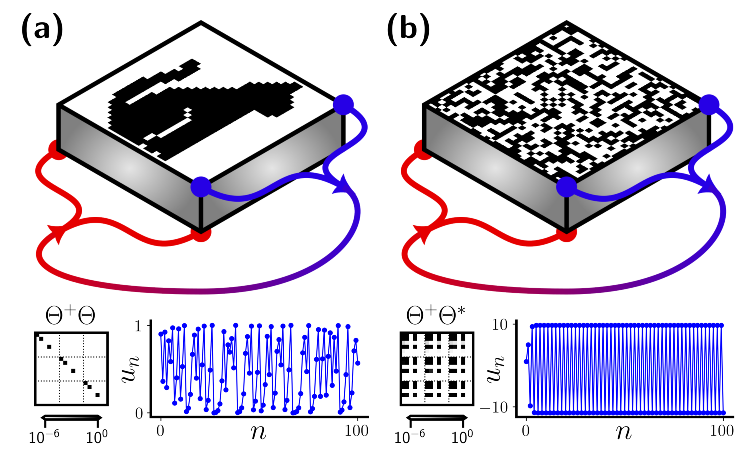}
    \caption{
        Computation of $u_n = 4 u_{n-1} (1-u_{n-1})$ conditioned on a specific memory state.
        (a) Memory state associated with the programming conditions defined in Fig.~\ref{fig:programming}.
        (b) Emergence of an untrained attractor for an irrelevant memory state.
        The degradation of computational functionality is quantified by the loss of diagonality in the matrix $\Theta^{+}\Theta^{*}$, where $\Theta^{*}\equiv\mathcal{R}^{\top}\mathcal{R}^*$ and $\mathcal{R}^*$ denotes the partial derivatives of the reservoir states evaluated at an irrelevant memory state.
    }
    \label{fig:memory}
\end{figure}

\textit{Programming using the universal approximation}.---
Figure~\ref{fig:programming} demonstrates the programming of our spintronic RC system in the regime $\tau \gg 1$.
Under the assumption that $\bm{H}_{i}^{(0)} \cdot \bm{H}_{i}^{(RC)} = 0$, this parameter condition corresponds to the UAP for mappings of the current input $f^{\rm id}(u_n)$
\cite{
    hornik1991approximation,pinkus1999approximation}.
Because the UAP ensures that $Wh \approx f^{\rm id}(u_n)$ for any continuous function $f^{\rm id}$, the analytic bases of $u_n$ must always be programmable for sufficiently large $N$; in the limit $\tau,N\rightarrow\infty$, the programmability matrix $\Theta^{+}\Theta$ becomes diagonal, providing functional components of $u_n$ and none of the past inputs [Fig.~\ref{fig:programming}(a)] (see Sec.~III~C in the SM). 
For example, trained artificial neural networks (ANNs) can be regarded as maps of the current input $f^{\rm id}(u_n)$, which can be emulated by programming in this regime.
In Fig.~\ref{fig:programming}(b), the spintronic RC performs Iris classification by emulating a trained ANN.
Furthermore, by keeping the variables $\mu$ symbolic in $\mathcal{S}$, the programming framework allows readouts to incorporate external control signals $\mu$ through $W(\mu) \equiv \mathcal{S}(\mu)\Theta^+\mathcal{R}^\top$.
When $\mu$ serves as a bifurcation parameter, an entire bifurcation structure can be programmed without decomposition into the analytic bases of $\mu$ [Fig.~\ref{fig:programming}(c)].
Figure~\ref{fig:programming}(d) shows the programming of a numerical solver for a fifth-order algebraic equation; here, the location of the RC attractor represents the solution for a given set of coefficients $\mu=\bm{a}$.
Data-driven optimization of the readout cannot easily achieve these types of computations, as they require explicit relationships between parameters $\mu$ and inputs $\{u_n\}$ to formulate the parameter dependencies in $W(\mu)$. 
Nonetheless, the impact of slight discrepancies between $f$ and $f^{\rm id}$ on the intended functionality must be considered in physical implementations (see Sec.~IV in the SM, which includes Refs.~\cite{san2000stochastic,garcia1998langevin, d2006midpoint}).

\textit{Conclusion.}---
In our framework, once the mathematical expression for the ideal computation $f^{\rm id}$ is obtained, the physical substrate can be replaced \cite{komatsu2024algebraic}:
Programming PRC transfers the prescribed knowledge to physical hardware, utilizing the underlying governing equations and decompositions $\{\partial^{k_d}_p\}$.
For example, conventional pseudorandom number generators running on silicon computers could be replaced by programmed chaotic dynamics in spintronic memory devices [Fig.~\ref{fig:memory}(a)] \cite{phatak1995logistic}.
Furthermore, since the source code $\mathcal{S}$ modifies only the readout component, arbitrary functionalities can be implemented using the same disassembly matrix $\mathcal{R}$ [Fig.~\ref{fig:programming}(b)--(d)].

Utilizing the governing equations of a PRC system satisfying conditions (i) and (ii), designers can derive the matrix $\Theta^+\Theta$ according to specific requirements $\{\partial^{k_d}_p\}$.
This matrix characterizes the consistency of these requirements within the system and depends intrinsically on system parameters.
In our spintronic RC system, the MRAM memory states $\{\bm{H}_{i}^{(0)}\}_{i=1}^{N}$ can also modulate $\Theta^+\Theta$;
consequently, programmed functionalities are directly linked to these states (see Fig.~\ref{fig:memory} for details).
This property may be beneficial for implementing password-protected computations or encryption schemes that exploit production errors \cite{gao2020physical}.

Designers may encounter off-diagonal or zero diagonal entries in $\Theta^+\Theta$ in their hardware architectures.
These entries represent a form of expressivity unique to PRC devices that is not observed in ANNs whose expressivity is guaranteed by the UAP [see Fig.~\ref{fig:Theta+Theta}(a) and SM].
Since the design requirements $\{\partial^{k_d}_p\}$ act as hyperparameters, the optimal choice of these requirements must be identified for each physical system based on the diagonality of $\Theta^+\Theta$.
However, this process may be subject to the curse of dimensionality as the input dimensionality increases.
Future work will focus on experimental validation, high-dimensional input analysis, and a rigorous mathematical treatment of the $\Theta^+\Theta$ matrix.

\begin{acknowledgments}
    Y. I. acknowledges support from JSPS KAKENHI Grant No. JP23KJ0331. K. N. acknowledges support from JSPS KAKENHI Grant Nos. JP24K21323 and JP25H01134 and from project JPNP14004, commissioned by the New Energy and Industrial Technology Development Organization (NEDO).
\end{acknowledgments}


\bibliography{PSR}

@PREAMBLE{
 "\providecommand{\noopsort}[1]{}" 
 # "\providecommand{\singleletter}[1]{#1}%" 
}

@article{jaeger2023toward,
  title={Toward a formal theory for computing machines made out of whatever physics offers},
  author={Jaeger, Herbert and Noheda, Beatriz and Van Der Wiel, Wilfred G},
  journal={Nature communications},
  volume={14},
  number={1},
  pages={4911},
  year={2023},
  publisher={Nature Publishing Group UK London}
}

@article{nakajima2020physical,
  doi = {10.35848/1347-4065/ab8d4f},
  url = {https://dx.doi.org/10.35848/1347-4065/ab8d4f},
  year = {2020},
  month = {may},
  publisher = {IOP Publishing},
  volume = {59},
  number = {6},
  pages = {060501},
  author = {Kohei Nakajima},
  title = {Physical reservoir computing—an introductory perspective},
  journal = {Japanese Journal of Applied Physics}
}

@book{nakajima2021reservoir,
  title={Reservoir Computing},
  subtitle={Theory, Physical Implementations, and Applications},
  author={Nakajima, Kohei and Fischer, Ingo},
  year={2021},
  publisher={Springer, Singapore}
}

@article{momeni2025training,
  title={Training of physical neural networks},
  author={Momeni, Ali and Rahmani, Babak and Scellier, Benjamin and Wright, Logan G and McMahon, Peter L and Wanjura, Clara C and Li, Yuhang and Skalli, Anas and Berloff, Natalia G and Onodera, Tatsuhiro and others},
  journal={Nature},
  volume={645},
  number={8079},
  pages={53--61},
  year={2025},
  publisher={Nature Publishing Group UK London}
}

@article{wright2022deep,
  title={Deep physical neural networks trained with backpropagation},
  author={Wright, Logan G and Onodera, Tatsuhiro and Stein, Martin M and Wang, Tianyu and Schachter, Darren T and Hu, Zoey and McMahon, Peter L},
  journal={Nature},
  volume={601},
  number={7894},
  pages={549--555},
  year={2022},
  publisher={Nature Publishing Group UK London}
}

@article{nakajima2022physical,
  title={Physical deep learning with biologically inspired training method: gradient-free approach for physical hardware},
  author={Nakajima, Mitsumasa and Inoue, Katsuma and Tanaka, Kenji and Kuniyoshi, Yasuo and Hashimoto, Toshikazu and Nakajima, Kohei},
  journal={Nature communications},
  volume={13},
  number={1},
  pages={7847},
  year={2022},
  publisher={Nature Publishing Group UK London}
}

@article{wang2026embodying,
  title={Embodying physical computing into soft robots},
  author={Wang, Jun and Zhou, Ziyang and Kahak, Ardalan and Li, Suyi},
  journal={Nature Communications},
  year={2026},
  publisher={Nature Publishing Group UK London}
}

@article{akashi2022coupled,
author = {Akashi, Nozomi and Kuniyoshi, Yasuo and Tsunegi, Sumito and Taniguchi, Tomohiro and Nishida, Mitsuhiro and Sakurai, Ryo and Wakao, Yasumichi and Kawashima, Kenji and Nakajima, Kohei},
title = {A Coupled Spintronics Neuromorphic Approach for High-Performance Reservoir Computing},
journal = {Advanced Intelligent Systems},
volume = {4},
number = {10},
pages = {2200123},
doi = {https://doi.org/10.1002/aisy.202200123},
url = {https://advanced.onlinelibrary.wiley.com/doi/abs/10.1002/aisy.202200123},
year = {2022}
}

@article{komatsu2024algebraic,
  title = {Algebraic design of physical computing system},
  journal = {Physica D: Nonlinear Phenomena},
  volume = {470},
  pages = {134382},
  year = {2024},
  issn = {0167-2789},
  doi = {https://doi.org/10.1016/j.physd.2024.134382},
  url = {https://www.sciencedirect.com/science/article/pii/S0167278924003324},
  author = {Mizuka Komatsu and Takaharu Yaguchi and Kohei Nakajima}
}

@ARTICLE{jaeger2007echo,
AUTHOR = {Jaeger, H. },
TITLE   = {{E}cho state network},
YEAR    = {2007},
JOURNAL = {Scholarpedia},
VOLUME  = {2},
NUMBER  = {9},
PAGES   = {2330},
DOI     = {10.4249/scholarpedia.2330},
NOTE    = {revision \#196567}
}

@article{hornik1991approximation,
  title = {Approximation capabilities of multilayer feedforward networks},
  journal = {Neural Networks},
  volume = {4},
  number = {2},
  pages = {251-257},
  year = {1991},
  issn = {0893-6080},
  doi = {https://doi.org/10.1016/0893-6080(91)90009-T},
  author = {Kurt Hornik},
}

@article{pinkus1999approximation,
  title={Approximation theory of the MLP model in neural networks},
  volume={8},
  DOI={10.1017/S0962492900002919},
  journal={Acta Numerica},
  author={Pinkus, Allan},
  year={1999},
  pages={143–195}
}

@article{chen2025spintronic,
  title={Spintronic reservoir computing with interpretable nonlinearity},
  author={Chen, Jiaxuan and Song, Yicheng and Hirose, Akira},
  journal={Physical Review Research},
  volume={7},
  number={1},
  pages={013310},
  year={2025},
  publisher={APS}
}

@article{gilbert2004phenomenological,
  title={A phenomenological theory of damping in ferromagnetic materials},
  author={Gilbert, Thomas L},
  journal={IEEE transactions on magnetics},
  volume={40},
  number={6},
  pages={3443--3449},
  year={2004},
  publisher={IEEE}
}

@article{akashi2020input,
  title={Input-driven bifurcations and information processing capacity in spintronics reservoirs},
  author={Akashi, Nozomi and Yamaguchi, Terufumi and Tsunegi, Sumito and Taniguchi, Tomohiro and Nishida, Mitsuhiro and Sakurai, Ryo and Wakao, Yasumichi and Nakajima, Kohei},
  journal={Physical Review Research},
  volume={2},
  number={4},
  pages={043303},
  year={2020},
  publisher={APS}
}

@article{imai2022noise,
  title={Noise-induced synchronization of spin-torque oscillators},
  author={Imai, Yusuke and Tsunegi, Sumito and Nakajima, Kohei and Taniguchi, Tomohiro},
  journal={Physical Review B},
  volume={105},
  number={22},
  pages={224407},
  year={2022},
  publisher={APS}
}

@article{imai2025gradient,
  title={Gradient-based optimization of spintronic devices},
  author={Imai, Yusuke and Liu, Shuhong and Akashi, Nozomi and Nakajima, Kohei},
  journal={Applied Physics Letters},
  volume={126},
  number={8},
  year={2025},
  publisher={AIP Publishing}
}

@Article{kurebayashi2026metrics,
author={Kurebayashi, Hidekazu
  and Finocchio, Giovanni
  and Everschor-Sitte, Karin
  and Gartside, Jack C.
  and Taniguchi, Tomohiro
  and Litvinenko, Artem
  and Kumar, Akash
  and {\AA}kerman, Johan
  and Vasilaki, Eleni
  and Sel{\c{c}}uk, Kemal
  and {\c{C}}amsar{\i}, Kerem Y.
  and Madhavan, Advait
  and Fukami, Shunsuke},
  title={Metrics for spin-based computing},
  journal={Nature Reviews Physics},
  year={2026},
  month={Apr},
  day={01},
  volume={8},
  number={4},
  pages={208-225},
  issn={2522-5820},
  doi={10.1038/s42254-025-00918-1},
}

@article{sebastian2020memory,
  title={Memory devices and applications for in-memory computing},
  author={Sebastian, Abu and Le Gallo, Manuel and Khaddam-Aljameh, Riduan and Eleftheriou, Evangelos},
  journal={Nature nanotechnology},
  volume={15},
  number={7},
  pages={529--544},
  year={2020},
  publisher={Nature Publishing Group UK London}
}

@article{sun2023full,
  title={A full spectrum of computing-in-memory technologies},
  author={Sun, Zhong and Kvatinsky, Shahar and Si, Xin and Mehonic, Adnan and Cai, Yimao and Huang, Ru},
  journal={Nature Electronics},
  volume={6},
  number={11},
  pages={823--835},
  year={2023},
  publisher={Nature Publishing Group UK London}
}

@article{marrows2024neuromorphic,
  title={Neuromorphic computing with spintronics},
  author={Marrows, Christopher H and Barker, Joseph and Moore, Thomas A and Moorsom, Timothy},
  journal={npj Spintronics},
  volume={2},
  number={1},
  pages={12},
  year={2024},
  publisher={Nature Publishing Group UK London}
}

@article{grigoryeva2019differentiable,
  author  = {Lyudmila Grigoryeva and Juan-Pablo Ortega},
  title   = {Differentiable reservoir computing},
  journal = {Journal of Machine Learning Research},
  year    = {2019},
  volume  = {20},
  number  = {179},
  pages   = {1--62},
  url     = {http://jmlr.org/papers/v20/19-150.html}
}

@article{kim2023neural,
	title = {A neural machine code and programming framework for the reservoir computer},
	volume = {5},
	issn = {2522-5839},
	url = {https://www.nature.com/articles/s42256-023-00668-8},
	doi = {10.1038/s42256-023-00668-8},
	number = {6},
	urldate = {2024-10-06},
	journal = {Nature Machine Intelligence},
	author = {Kim, Jason Z. and Bassett, Dani S.},
	year = {2023},
	pages = {622--630},
}

@article{phatak1995logistic,
  title = {Logistic map: A possible random-number generator},
  author = {Phatak, S. C. and Rao, S. Suresh},
  journal = {Phys. Rev. E},
  volume = {51},
  issue = {4},
  pages = {3670--3678},
  numpages = {0},
  year = {1995},
  month = {Apr},
  publisher = {American Physical Society},
  doi = {10.1103/PhysRevE.51.3670},
  url = {https://link.aps.org/doi/10.1103/PhysRevE.51.3670}
}

@article{gao2020physical,
  title={Physical unclonable functions},
  author={Gao, Yansong and Al-Sarawi, Said F and Abbott, Derek},
  journal={Nature Electronics},
  volume={3},
  number={2},
  pages={81--91},
  year={2020},
  publisher={Nature Publishing Group UK London}
}

@article{d2006midpoint,
    author = {d’Aquino, M. and Serpico, C. and Coppola, G. and Mayergoyz, I. D. and Bertotti, G.},
    title = {Midpoint numerical technique for stochastic Landau-Lifshitz-Gilbert dynamics},
    journal = {Journal of Applied Physics},
    volume = {99},
    number = {8},
    pages = {08B905},
    year = {2006},
    month = {04},
    issn = {0021-8979},
    doi = {10.1063/1.2169472},
    url = {https://doi.org/10.1063/1.2169472},
}

@incollection{san2000stochastic,
  author = {San Miguel, Maxi and Toral, Ra{\'u}l},
  title = {Stochastic Effects in Physical Systems},
  booktitle = {Instabilities and Nonequilibrium Structures VI},
  editor = {Tirapegui, Enrique and Mart{\'i}nez, Javier and Tiemann, Rolando},
  year = {2000},
  publisher = {Springer Dordrecht},
  series = {Nonlinear Phenomena and Complex Systems},
  volume = {5},
  pages = {35--127},
  isbn = {978-0-7923-6129-9},
  doi = {10.1007/978-94-011-4247-2_2},
  url = {https://doi.org/10.1007/978-94-011-4247-2_2}
}

@article{garcia1998langevin,
  title = {Langevin-dynamics study of the dynamical properties of small magnetic particles},
  author = {Garc\'{\i}a-Palacios, Jos\'e Luis and L\'azaro, Francisco J.},
  journal = {Phys. Rev. B},
  volume = {58},
  issue = {22},
  pages = {14937--14958},
  numpages = {0},
  year = {1998},
  month = {Dec},
  publisher = {American Physical Society},
  doi = {10.1103/PhysRevB.58.14937},
  url = {https://link.aps.org/doi/10.1103/PhysRevB.58.14937}
}

\end{document}